\documentclass[%
 reprint,
superscriptaddress,
showpacs,preprintnumbers,
nofootinbib,
nobibnotes,
 amsmath,amssymb,
aps,
pra,
floatfix,
]{revtex4-1}

\usepackage{graphicx}
\usepackage{dcolumn}
\usepackage{bm}
\usepackage{hyperref}
\usepackage{tabularx}
\usepackage{amsmath}

\usepackage{mwe}    
\usepackage{subfloat}

\usepackage[normalem]{ulem}

\usepackage[utf8]{inputenc}

\usepackage{natbib}

\usepackage{color}
\usepackage{soul}

\usepackage{breqn}

\begin{document}

\title{Pressureless stationary solutions in a Newton-Yukawa gravity model}

\author{Tiago D. Ferreira}
\email{Corresponding author: tiagodsferreira@hotmail.com}
\affiliation{Departamento de Física e Astronomia da Faculdade de Ciências da Universidade do Porto$,$  Rua do Campo Alegre 687$,$ 4169-007 Porto$,$ Portugal}
\affiliation{INESC TEC$,$ Centre of Applied Photonics$,$ Rua do Campo Alegre 687$,$ 4169-007 Porto$,$ Portugal}

\author{João Novo}
\affiliation{Departamento de Física e Astronomia da Faculdade de Ciências da Universidade do Porto$,$ Rua do Campo Alegre 687$,$ 4169-007 Porto$,$ Portugal}
\affiliation{Centro de Física das Universidades do Minho e do Porto$,$ Rua do Campo Alegre 687$,$ 4169-007$,$ Porto$,$ Portugal}

\author{Nuno A. Silva}
\affiliation{Departamento de Física e Astronomia da Faculdade de Ciências da Universidade do Porto$,$ Rua do Campo Alegre 687$,$ 4169-007 Porto$,$ Portugal}
\affiliation{INESC TEC$,$ Centre of Applied Photonics$,$ Rua do Campo Alegre 687$,$ 4169-007 Porto$,$ Portugal}

\author{A. Guerreiro}
\affiliation{Departamento de Física e Astronomia da Faculdade de Ciências da Universidade do Porto$,$ Rua do Campo Alegre 687$,$ 4169-007 Porto$,$ Portugal}
\affiliation{INESC TEC$,$ Centre of Applied Photonics$,$ Rua do Campo Alegre 687$,$ 4169-007 Porto$,$ Portugal}

\author{O. Bertolami}
\affiliation{Departamento de Física e Astronomia da Faculdade de Ciências da Universidade do Porto$,$ Rua do Campo Alegre 687$,$ 4169-007 Porto$,$ Portugal}
\affiliation{Centro de Física das Universidades do Minho e do Porto$,$ Rua do Campo Alegre 687$,$ 4169-007$,$ Porto$,$ Portugal}

\date{\today}

\begin{abstract}

Non-minimally coupled curvature-matter gravity models are an interesting alternative to the Theory of General Relativity and to address the dark energy and dark matter cosmological problems. These models have complex field equations that prevent a full analytical study. Nonetheless, in a particular limit, the behavior of a matter distribution can, in these models, be described by a Schr\"odinger-Newton system. In nonlinear optics, the Schr\"odinger-Newton system can be used to tackle a wide variety of relevant situations and several numerical tools have been developed for this purpose. Interestingly, these methods can be adapted to study General Relativity problems as well as its extensions. In this work, we report the use of these numerical tools to study a particular non-minimal coupling model that introduces two new potentials, an attractive Yukawa potential and a repulsive potential proportional to the energy density. Using the imaginary-time propagation method we have shown that stationary solutions arise even at low energy density regimes.


\end{abstract}

\maketitle


\section{\label{sec:level1}Introduction} 

In recent years various alternative theories of gravity have been proposed to extend General Relativity (GR) and studied to address some well-known cosmological difficulties such as dark energy and dark matter. Non-minimally coupled (NMC) curvature-matter gravity models \cite{Bertolami:2007gv} were proposed to approach these problems under a different perspective, but are also particularly interesting as they give rise to a rich lore of features and have a wide range of astrophysical and cosmological implications (see, for instance, Ref. \cite{OBParamos2014}, for a review). NMC curvature-matter gravity models extend GR and the so-called $f(R)$ models of gravity by allowing for an extra curvature coupling to the matter Lagrangian density. In its most general setting, NMC curvature-matter gravity models are characterized by two functions of the scalar curvature, $f_{1,2}(R)$. Function $f_1(R)$ allows for going beyond the linear scalar curvature term of the Einstein-Hilbert action whereas $f_2(R)$ generalizes the minimal coupling between matter and geometry that involves the square root of the positive valued determinant of the metric, $\sqrt{|g|}$, so to keep the element of volume invariant, and covariant derivatives. The non-minimal coupling gives rise to very convoluted field equations that cannot, in general, be treated analytically and poses challenges to the existing numerical methods. However, as previously shown \cite{PhysRevE.101.023301}, under the right assumptions, the field equations for specific matter distributions can be transformed into a Schrödinger-Newton system of equations.

The Schrödinger-Newton model, also known in the literature as Schrödinger-Poisson, is commonly used in nonlinear optics for describing light propagating in nonlinear and nonlocal media, under the paraxial approximation \cite{cubic_quintic_material,NLC_playground,boson_stars,supress_Drag_force_nature}, and self-gravitating Bose-Einstein condensates \cite{solutions_principal}, among other situations \cite{SN_foundations,pethick_smith_2008,quantum_fluids_of_light}. Due to the wide applicability of this model, many numerical schemes were developed \cite{SSFM_NSE}, with the Split-Step Fourier method (SSFM) being the most suitable one. This numerical scheme, apart from being easy to implement, also allows searching for the existence of stationary solutions through the imaginary-time propagation method \cite{PhysRevE.62.7438,Lehtovaara2007}, whereby propagating an initial random ansatz the system ultimately converges into a stationary solution, if it exists. When both these schemes are applied to these new theories of gravity, they open the possibility to study the dynamics imposed by the model settings and search for the existence of new stationary configurations simply and efficiently. Thus, this set of techniques stand as a promising tool to probe these new gravity models, as they can be used to quickly test and impose constraints on the model features. This idea was previously explored by our research group for NMC models \cite{PhysRevE.101.023301}, however in this work we focus on a different approximation regime and consider the imaginary-time propagation method to search for stationary solutions.

Besides the computational advantages, the Schrödinger-Newton approach is also promising as it can be used to develop table-top experiments that, under certain conditions, can emulate the dynamical features of some cosmological systems. These phenomena range from gravitational effects \cite{gravitational_effects}, boson stars \cite{boson_stars}, scalar dark-matter models \cite{dark_matter,Paredes2020}, false-vacuum decay \cite{Fialko2015,Braden2019a}, to acoustic black-holes \cite{Faccio2012a,PhysRevA.78.063804,Vocke2018} to study the formation of Hawking-radiation \cite{Drori2019,Nova2019}, superradiance \cite{2020_penrose_superradiance, PhysRevD100024037}, the Penrose effect \cite{Solnyshkov2019} and even the formation of scalar clouds \cite{PhysRevD.103.045004}. At the conceptual level, these so-called optical analogue experiments allow to test features of the theoretical models in an experimental setting, gaining new insights through interdisciplinary research. Leveraging this cross-fertilization, the alternative theories of gravity explored in this work stand as an interesting candidate for further development and implementation of a new class of analogues dedicated to explore their dynamics under controlled experimental conditions.

In this work, we make use of advanced high-performance computing tools previously developed at our research group in the context of nonlinear optics \cite{meu_artigo, Ferreira2019, artigo_nuno}, to detect and to explore how NMC curvature-matter models of gravity give rise to stationary large scale distributions of mass. In particular, we focus on a particular non-minimal coupled curvature-matter gravity model, where functions $f_1(R)$ and $f_2(R)$ are expanded up to the second and first order in the curvature $R$, respectively, and assume that matter at the relevant scales behaves as a fluid. These assumptions allow for describing the system by a set of hydrodynamic equations, and two potentials arise in this context, an attractive Yukawa potential besides the Newtonian one, and a repulsive potential proportional to the energy density. Following this procedure, the Schrödinger-Newton model is obtained through the application of a Madelung transformation \cite{madelung}, and, by considering the Thomas-Fermi approximation, we show that this gravity model supports stationary solutions in the absence of a pressure term. We then calculate these solutions through the imaginary-time propagation method and compare the results with some analytical predictions. Finally, we discuss the implications of these pressureless solutions in the context of these alternative gravity models.

\section{\label{sec:level2}Gravitational Model}
In this work we will focus on a previously proposed NMC curvature-matter gravity model \cite{Bertolami:2007gv} that admits two functions of the scalar curvature, $f_1(R)$, and, $f_2(R)$. The physical model is then described by the following action:
\begin{equation}
S=\int\mathrm{d}^4x \sqrt{|g|} \left[\frac{1}{2}f_1(R)+\left(1+f_2(R)\right)\mathcal{L}_m\right]\label{NMCaction}\,,
\end{equation}
where $\mathcal{L}_m$ is the matter Lagrangian density, with GR trivially recovered for the choice:
\begin{equation}
f_1(R)=\frac{R}{\kappa}\,,\quad f_2(R)=0\,.
\end{equation}
In the metric formalism, the field equations for this model are:
\begin{equation}
\begin{split}
\left(F_1+2F_2\mathcal{L}_m\right)&R_{\mu\nu}-\frac{f_1}{2}g_{\mu\nu}=\left(1+f_2\right)T_{\mu\nu}\\
&+\left(\nabla_\mu\nabla_\nu-g_{\mu\nu}\square\right)\left(F_1+2F_2\mathcal{L}_m\right)\,,\label{FieldEquations}
\end{split}
\end{equation}
where $F_i\equiv\partial f_i/\partial R$, and $\Box$ is the D'Alembertian operator defined as $\Box=g^{\mu\nu}\nabla_\mu\nabla_\nu$ in terms of covariant derivatives from a Levi-Civita connection.

From these equations it is possible to deduce one of the most distinctive features of the NMC curvature-matter gravity, namely that the energy-momentum tensor is not covariantly conserved. In fact, taking the covariant derivative of Eq. (\ref{FieldEquations}) we get:
\begin{equation}
\nabla_\mu T^{\mu\nu}=\frac{F_2}{1+f_2}\left(g^{\mu\nu}\mathcal{L}_m-T^{\mu\nu}\right)\nabla_\mu R\,. \label{covariantderivativeNMC}
\end{equation}

Additional implications of the model have been extensively discussed in the literature and can be found in previously works of one of the authors \cite{OBParamos2014,OBLoboParamos2008,OBGomesJCAP2014,OBGomesJCAP2017,OBGomesEPL2018,OBParamos2020,OBGomes2020}. 
Considering the functions $f_i(R),\,i=1,2$ to be analytical around $R=0$, we can make use of a Taylor expansion and express
\begin{equation}
\begin{split}
f_1(R)&=\frac{1}{\kappa}(a_1 R+a_2 R^2)+\mathcal{O}(R^3)\,,\\
f_2(R)&=q_1R+\mathcal{O}(R^2)\,.\label{NMCfunctions}
\end{split}
\end{equation}
This expansion have been examined in the context of oceanic experiments \cite{OceanExperiment} and consistency with the well known Cassini experiment is currently under scrutiny \cite{March2021}. Here, we will extend beyond these scales and explore it at an astrophysical context.

Taking the nonrelativistic limit, the solution of the field equations gives for the $00$ component of the metric \cite{1/cexpansion}
\begin{equation}
    g_{00}=-1+\frac{2}{c^2}\left[U +\frac{1}{3}\left(1-\frac{q_1}{a_2}\right)Y\right]+\mathcal{O}(c^{-4})\,
\end{equation}
containing both a Newtonian and a Yukawa potentials, defined through the equations
\begin{equation}
\nabla^2U=-4\pi G\rho,
\label{NewtPot}
\end{equation}
\begin{equation}
\left(\nabla^2-\frac{1}{\lambda^2}\right)Y=-4\pi G\rho\,,
\label{YukPot}
\end{equation}respectively. In this regime, the resulting hydrodynamic equations for a fluid with energy density, $\rho$, isotropic pressure, $P$, and velocity, $\vec{v}$, are \cite{OceanExperiment}
\begin{equation}
\frac{\partial\rho}{\partial t}+\nabla(\rho\Vec{v})=0
\label{eq:cosmo_hydro_1}
\end{equation}
\begin{equation}
\frac{\partial\Vec{v}}{\partial t}+(\Vec{v}\cdot\nabla)\Vec{v}=\nabla\left[U+\alpha_0Y-V_p-\frac{4\pi}{3}G\lambda^2\theta^2\rho\right],
\label{eq:cosmo_hydro_2}
\end{equation}
where $\theta=q_1/a_2$, $\alpha_0=(1-\theta)^2/3$ and $\lambda=\sqrt{6a_2}$. For the pressure potential, $V_p$, we will assume a polytropic relation between $\rho$ and $P$ given by $P=w\rho^n$, with $w$ and $n$ being constants chosen for each physical situation, which gives
\begin{equation}
V_p=\begin{cases}
w\ln(\rho) & n=1\\
\frac{nw}{n-1}\rho^{n-1} & n>1 \lor n<0
\end{cases}\,.
\end{equation}

Comparing Eq. (\ref{eq:cosmo_hydro_2}) with the ones obtained from the Newtonian limit of GR it is clear that two additional terms appear: the Yukawa potential, and a term depending on the energy density, with a proportionality coefficient $4\pi G \lambda^2 \theta^2/3$. In the subsequent sections we will explore some of the consequences of these terms.

Before advancing we point out that an additional assumption $a_2>0$ is required so that the range of the Yukawa potential is real and this is related to the Dolgov-Kawasaki stability criterion \cite{EnergyConditionsNMC,Nojiri_2003}:
\begin{equation}
    f_1''(R)+2\mathcal{L}_m f_2''(R)\geq 0\,.
\end{equation}
This requirement stems from the fact that the field equations for NMC curvature-matter gravity are greater than second order. Thus, the Ricci scalar is not algebraically related to the trace of the energy-momentum tensor but rather given by a differential equation. This means that $R$ is, in fact, a dynamical field and the Dolgov-Kawasaki stability criterion corresponds to the requirement that the field effective mass ($1/\lambda)$ is positive.

\section{\label{sec:level3}A cosmology-like setting}

Describing the gravitational system as a fluid characterized by the set of hydrodynamic Eqs. (\ref{eq:cosmo_hydro_1}) and (\ref{eq:cosmo_hydro_2}), is quite useful to explore the dynamics of the model as, by performing a Madelung transformation \cite{madelung}, the set of equations are transformed into a Schrödinger-Newton-Yukawa model \cite{gravity_hidro}. This transformation is achieved by writing $\psi=\sqrt{\rho}e^{i\Phi/\nu}$, where $\nu$ is an adjustable parameter with the same dimensions as the velocity field $\Phi$, such that $\rho=|\psi|^2$ and $\vec{v}=\nabla\Phi$ are the density and velocity of the fluid, respectively. By substituting this into the hydrodynamic equations, re-scaling the field $\psi=\frac{\psi}{\sqrt{\rho_0}}$ and normalizing the equations, it is straightforward to show that
\begin{equation}
\begin{split}
i\frac{\partial\psi}{\partial t}=-\frac{1}{2}\nabla^2\psi-\left[ U+\alpha_0Y-\frac{\theta^2\Gamma^2}{3}|\psi|^2\right]\psi
\\
+\gamma_0\frac{n}{n-1}|\psi|^{2(n-1)}\psi+V_B\psi
\end{split}
\label{eq:SN}
\end{equation}
where
\begin{align}
\nabla^2U&=-|\psi|^2,
\label{eq:newton}
\\
\left(\nabla^2-\frac{1}{\Gamma^2}\right)Y&=-|\psi|^2,
\label{eq:yukawa}
\end{align}
$\Gamma$ is the normalized $\lambda$ parameter, and $\gamma_0=\omega\rho_0^{n-3/2}/(\nu\sqrt{4\pi G})$ is a constant that depends on the polytropic exponent $n$ and measures the strength of the pressure term. The last term in the  Schrödinger equation is the well known Bohm quantum potential or quantum pressure, $V_B=\nabla^2\sqrt{\rho}/(2\sqrt{\rho})$. This potential can be removed from equation (\ref{eq:SN}), but it shows up instead in the hydrodynamic equations like a pressure gradient. This potential has no equivalent in classical fluids, and it is usually found when attempting to describe quantum mechanics from a hydrodynamic perspective or in nonlinear optics due to the diffraction. In the present context we can estimate the effect of this potential by considering $\rho$ to be given by a Gaussian distribution with width $R$, and it is the straightforward to show that $V_B\sim1/R^2$. Thus, the Bohm potential is only important in circumstances where the energy density $\rho$ varies greatly over the scale of interest ($\sim1$), that is in collapse situations. However, these scenarios will not be considered in this work, and thus the Bohm potential can be safely disregarded. Finally, it is pertinent to point out that, despite the use of some quantum mechanics features, it is not our aim to describe our model from a quantum perspective, but rather use the mathematical procedure provided by the Madelung transformation \cite{madelung}, which allows transforming the complete Vlasov-Poisson system into a system of equations that are easier to work with.

Before advancing we note that equations (\ref{eq:SN}), (\ref{eq:newton}) and (\ref{eq:yukawa}) admit, in specific contexts, self similar solutions, described by the following scaling transformation
\begin{align}
\vec{r}\rightarrow\vec{r}'&=\lambda_S^{-1}\vec{r}
\\
t\rightarrow t'&=\lambda_S^{-2} t
\\
\psi\left(\vec{r},t\right)\rightarrow \psi' \left(\vec{r}',t'\right)&=\lambda_S^2\psi\left(\lambda_S\vec{r}',\lambda_S^2t'\right),
\label{eq:scale}
\end{align}
with $\Gamma'=\lambda_S^{-1}\Gamma$. Indeed, it is easy to show that these relations are valid when the pressure term is negligible or when the polytropic exponent $n=3/2$. Thus, the solutions found under these assumptions are not restricted to the original setting of the system, but can be transformed to describe other systems at different spatial scales. Furthermore, this is also important for testing the validity of the gravitational model, since it is possible to search for scaling factors that correctly scale a certain solution that fits some particular observational data. This fact will be explored later when we focus on the pressureless regime.

\subsection{Numerical Methods}

The SSFM is commonly regarded as one of the most suitable schemes for the study of the Schrödinger-Newton-Yukawa system \cite{SSFM_NSE}, balancing accuracy with performance. It consists in integrating the system of equations in small steps by transforming the field between the reciprocal and direct spaces, as the kinetic term is better integrated in the first space, whereas the nonlinear potentials in the former. Besides yielding accurate results, this implementation allows to takes advantage from parallel computing techniques and exploit high throughput hardware such as Graphical Processing Units(GPU) for performing complex simulations in reasonable spans of time. For more details about the numerical scheme and its implementation see Refs. \cite{Ferreira2019, PhysRevE.101.023301}, and references therein.

\begin{figure*}[!ht]
	\begin{center}
		\includegraphics[width=0.8\textwidth]{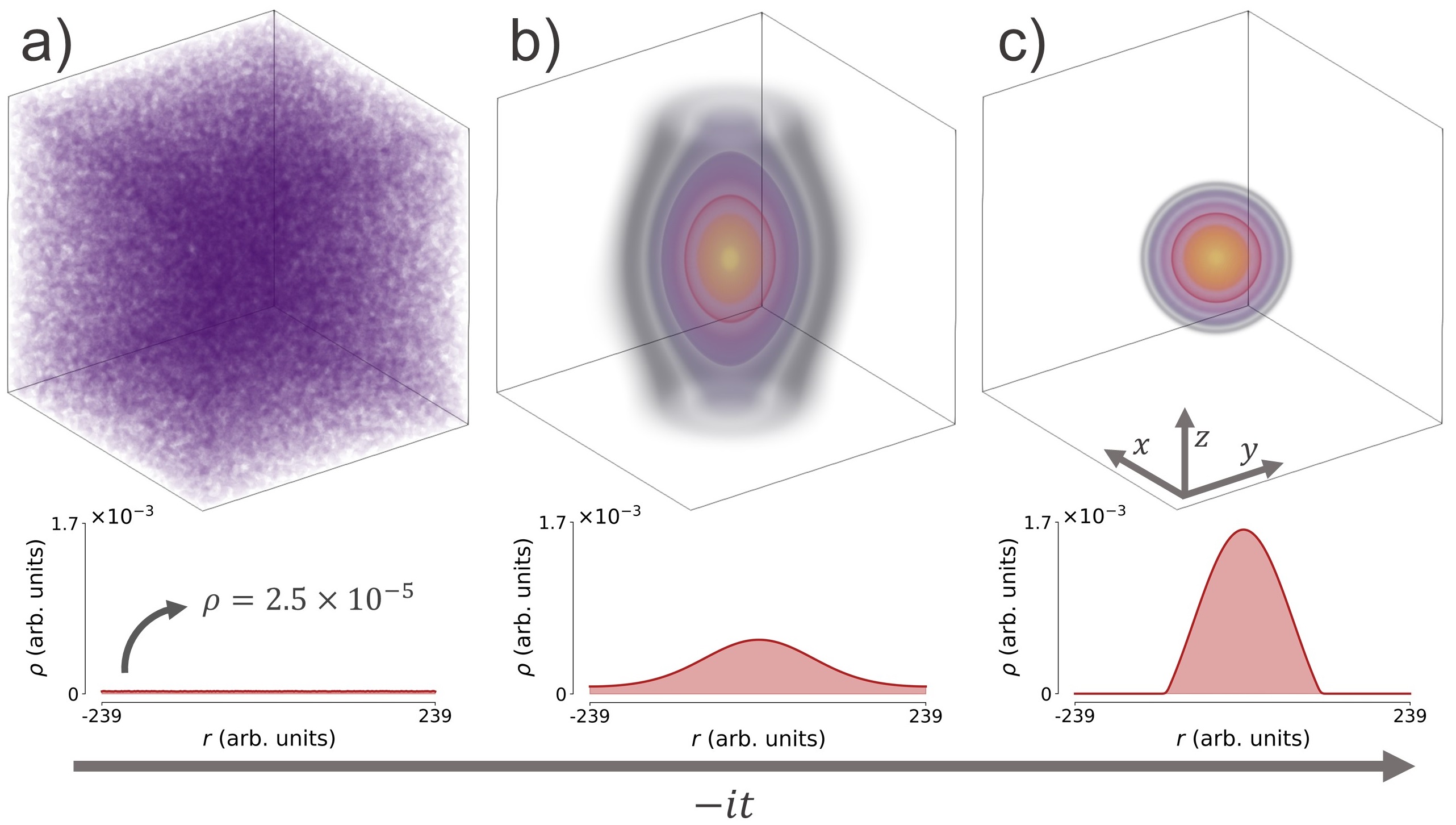}
		\caption{Tree-dimensional representation of the Imaginary-time propagation method for a solution with $\Gamma=60$ and $\alpha_0=1\times10^{-10}$. An initial random ansatz, given by Eq. (\ref{eq:initial_solution}) and corresponding to a), is propagated in an imaginary time scheme, and after some time the initial field converges into a stationary solution. b) The solution after $\sim119750$ iterations ($t\sim0.54$ Byr). c) The final solution after $\sim497500$ iterations ($t\sim2.18$ Byr). The bottom slices are cuts along the $z$-axis. The solutions are calculated in a tree-dimensional grid with $[256\times256\times256]$ points. In this figure and in the following results, in order to convert from the normalized into the physical units, it was considered $\rho_0=1\times10^{-20}kgm^{-3}$ and $\nu=1m^2s^{-1}$.}
		\label{fig:ITPM}
	\end{center}	
\end{figure*}

The most straightforward use of the method applied to this system is to explore the dynamics of energy distributions governed by the alternative gravity model and to examine the impact of its parameters. However, an interesting feature is that the method can also be used as a tool to search for stationary solutions. The technique for this purpose is known as the Imaginary-time propagation method (ITP) and consists in performing a Wick rotation on the time parameter $t\rightarrow-it$ \cite{PhysRevE.62.7438,Lehtovaara2007}. This transformation converts Eq. (\ref{eq:SN}) into a diffusion-like equation with emission and absorption terms. By propagating an initial random ansatz and ensuring that the total mass is conserved, the system converges to its ground state as $t\rightarrow\infty$, see Figure \ref{fig:ITPM}. As convergence criteria we choose the variation in the energy between integration steps to be below a certain threshold, where the energy is calculated from the Hamiltonian of the system
\begin{equation}
\begin{split}
\mathcal{H}=\frac{1}{2}|\nabla\psi|^2+\frac{\theta^2\Gamma^2}{6}|\psi|^4+\frac{\gamma_0}{n-1}|\psi|^{2n}
\\
 -\frac{1}{2}U|\psi|^2-\frac{\alpha_0}{2}Y|\psi|^2.
\end{split}
\end{equation}

The SSFM complemented with the ITPM is a robust and stable numerical scheme to explore the existence of stationary solutions. Still, depending on the parameters of the system, it can become numerically unstable. In particular, if we neglect the Bohm potential and the pressure term, the most critical term is the term proportional to the energy density, since in order to ensure the stability of the numerical scheme \cite{PhysRevE.62.7438}, we have to ensure that
\begin{equation}
\max\left(\left|\frac{\Gamma^2\theta^2}{3}|\psi|^2\Delta t\right|\right)<1,
\end{equation}
where $\Delta t$ is the integration step. This imposes limits on the numerical model, since, for large values of $\Gamma$, very small integration steps are required, which, on its turn, can increase the calculations time up to days or even months. On the other hand, we could also decrease the value of the normalized energy density, however, this would require values with limited physical interest. Having this term well controlled, the remaining ones are shown to be stable. Thus, although the numerical scheme does not impose a restrictive limit on the parameters that can be used, we are limited by the available time to compute the solutions and ultimately by the precision of the machine.

\section{Stationary Solutions}
The existence of stationary solutions requires a balance between the attractive and repulsive potentials which can be easily identified in the gravitational model considered in this work. On one hand, the potentials $U$ and $Y$ correspond to the attractive ones, and in the absence of counteracting forces, they force the system into a collapse. On the other hand, the remaining potentials, namely the pressure and the term proportional to the energy density, will force the system to expand. In general, the repulsion that balances the collapse is maintained by a collaboration between these two potentials. However, for systems where the pressure term can be neglected, it is interesting to notice that the current model may still allow the existence of solutions if solely sustained by the term proportional to the energy density. These stationary solutions, which will be investigated in the subsequent sections, are specific of the model under consideration and found no counterpart in GR nor in previously explored NMC curvature-matter gravity models \cite{PhysRevE.101.023301}.

\subsection{Analytical Approach}
To investigate the possibility of such stationary solutions we first explore the systems with an analytical approach, starting by recalling the hydrodynamic equation (\ref{eq:cosmo_hydro_2}) with the Bohm potential. We assume that we are in an astrophysical scenario where we can neglect the pressure term (pressureless fluid) as well as the Yukawa potential, motivated by the fact that the $\alpha_0$ parameter shall be negligible according to known observational bounds \cite{alpha_restrictions}. In a stationary regime, where $\partial_t\psi\rightarrow0$ and $\vec{v}=0$, the hydrodynamic Eq. (\ref{eq:cosmo_hydro_2}) can be rewritten, in dimensionless units, as
\begin{equation}
\rho+\frac{\theta^2\Gamma^2}{3}\nabla^2\rho-\frac{1}{2}\nabla^2\left(\frac{\nabla^2\sqrt{\rho}}{\sqrt{\rho}}\right)=0.
\label{eq:equilibrium_hydro_2}
\end{equation}
Since we are interested in solutions at large scales, the Bohm potential can be safely neglected, a simplification that corresponds to the common Thomas-Fermi (TF) approximation. Alternately, this approximation can be interpreted as if the kinetic term in Schrödinger Eq. (\ref{eq:SN}) could be neglected, since at large scales the contribution from this term to the total energy can be disregarded. Thus, in this approximation we have that
\begin{align}
\rho+\frac{\theta^2\Gamma^2}{3}\nabla^2\rho=0,
\label{eq:equilibrium_hydro_4}
\end{align}
which has an exact solution \cite{landen_solutions,solutions_principal} given by
\begin{equation}
\rho(r)=\begin{cases}
\frac{\rho_M\mathcal{R}}{\pi r}\sin\left(\frac{\pi r}{\mathcal{R}}\right) & r\leq \mathcal{R}\\
0 & r>\mathcal{R}
\end{cases}\,,
\label{eq:solution}
\end{equation}
where the maximum density $\rho_M$ can be obtained in terms of the total mass of the system, $M$, by integration of the solution over the volume as
\begin{equation}
\rho_M=\frac{\pi M}{4\mathcal{R}^3},
\label{eq:rho_0}
\end{equation}and the radius of the solution $\mathcal{R}$ at which the energy density vanishes (compact support) is
\begin{equation}
\mathcal{R}=\frac{\pi\theta\Gamma}{\sqrt{3}}.
\label{eq:radius}
\end{equation}
It is interesting to notice that this value only depends on the gravitational parameters of the model, $\theta$ and $\Gamma$, and is independent of the total mass of the system. However, we need to be careful since both, too large or too small masses, can violate the assumptions used for the TF approximation. In the first scenario, systems of larger mass are associated with an higher density. In this case, the pressureless regime is no longer valid, breaking the initial assumption of  Eq. (\ref{eq:equilibrium_hydro_4}). On the opposite side, the small mass limit is associated with an ultra low-density of the system for which the potential proportional to the energy density can become of the same order of magnitude as the Bohm potential. In such case, which happens when
\begin{equation}
\frac{1}{\mathcal{R}^{2}}\sim\frac{\theta^{2}\Gamma^{2}}{3}\frac{M}{\mathcal{R}^{3}}\Rightarrow M\sim\frac{\pi\sqrt{3}}{\theta\Gamma},
\end{equation}
the contributions of both will be of the same order which also invalidates the TF approximation.

\subsection{\label{sec:level4}Numerical Results}
\begin{figure}[!ht]
	\begin{center}
		\includegraphics[width=0.35\textwidth]{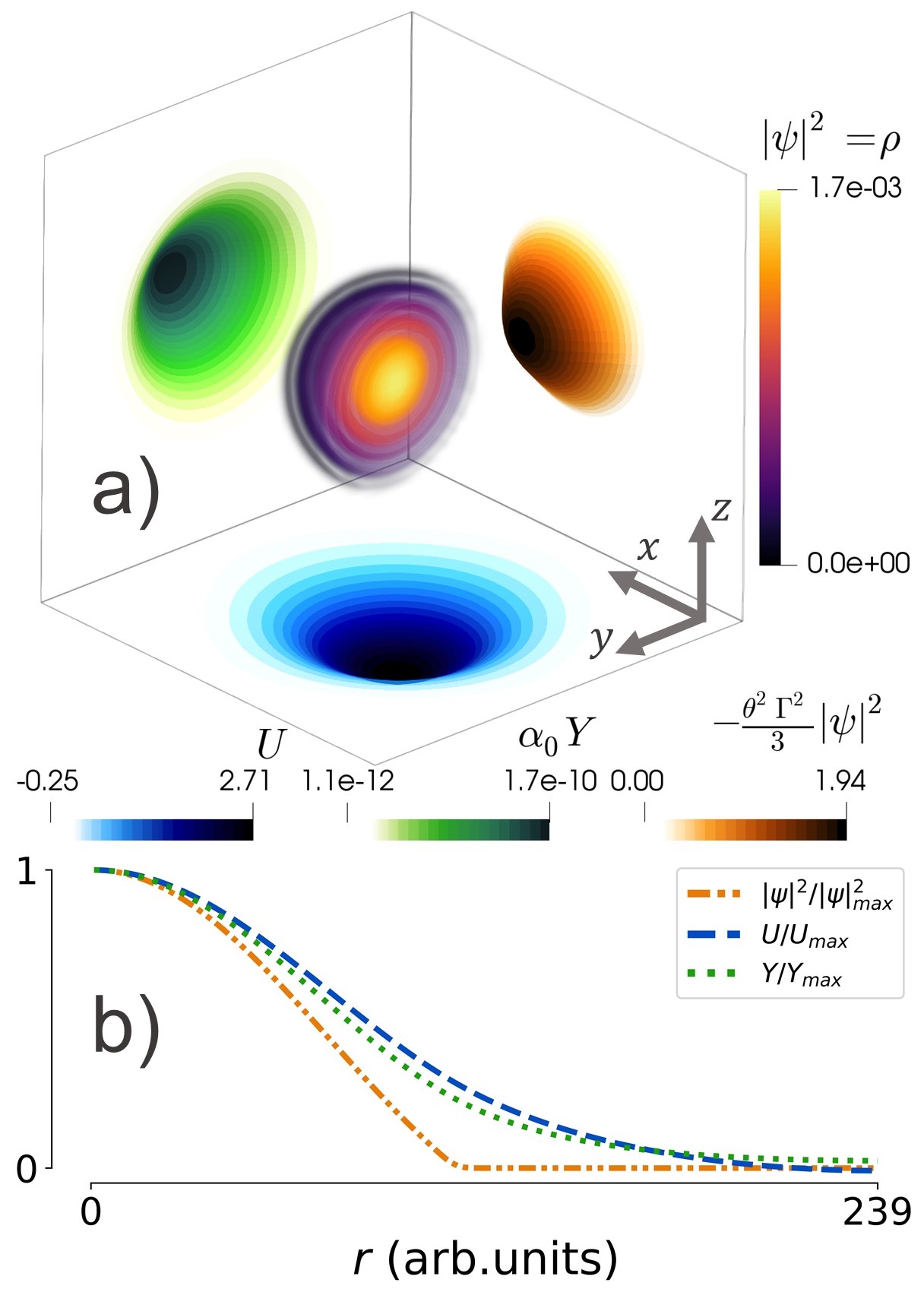}
		\caption{Representation of a stationary solution with $\Gamma=60$ and $\alpha_0=1\times10^{-10}$. a) 3D representation of the stationary solution and the potential's profiles. The color scales are in arbitrary units. b) Profile comparison of the potentials.}
		\label{fig:solution_3D}
	\end{center}	
\end{figure}

Let us now numerically examine the stationary solutions discussed in the previous section with the imaginary-time propagation method depicted in Figure \ref{fig:ITPM}. To avoid inducing symmetries in the profile of the solution, we started the simulations with a uniform field of amplitude $A$ destabilized by a small random noise $\epsilon(\vec{r})$
\begin{equation}
\psi(\vec{r},t=0)=A+\epsilon(\vec{r}).
\label{eq:initial_solution}
\end{equation}
This initial state is propagated under the ITP method and the system ultimately converges towards a stationary solution. In Figure \ref{fig:solution_3D} a particular solution is shown in a 3D representation, as well as the relevant potentials. It is easily seen that the solution is stabilized solely by the potential proportional to the energy density and the Newtonian potential, since the contribution from the Yukawa potential is negligible. Furthermore, from Figure \ref{fig:solution_3D}-b), where normalized slices of the potentials are plotted, the Newtonian and the Yukawa potentials are very similar. This happens because we have to consider large values for the $\Gamma$ parameter in order to satisfy the validity of TF approximation and, in this situation, the Yukawa potential resembles a Newtonian one. Thus, the stationary solutions found in the previous section remain valid even for large values of $\alpha_0$, and the effect of this extra nonlocal term can be considered in the hydrodynamic equations as $U\rightarrow(1+\alpha_0)U$, since $\alpha_0Y/\Gamma^2\sim0$.

\begin{figure}[!ht]
	\begin{center}
		\includegraphics[width=0.48\textwidth]{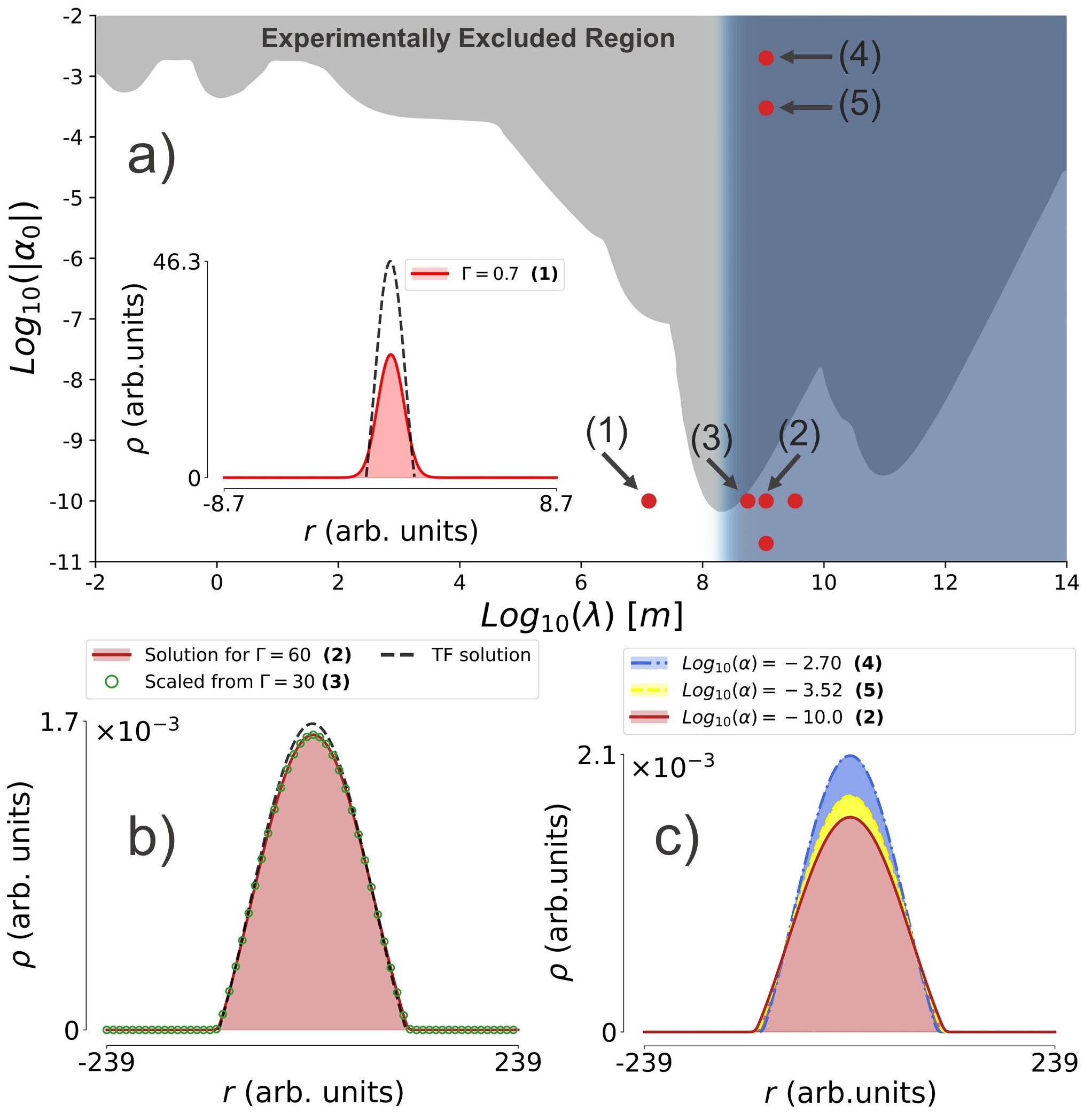}
		\caption{Stationary solutions found by the IMTP. a) Shows the regions prohibited for $\alpha_0$ as a function of the parameter $\lambda$, accordingly to Ref. \cite{alpha_restrictions}, and the position of the calculated solutions. The inset shows a solution outside the region of validity of the TF approximation. b) Stationary solution compared with the analytical solution given by Eq. (\ref{eq:solution}) and with a scaled one from a solution with $\Gamma=30$, corresponding to a scaling factor $\lambda_S=1/2$. c) Shows the impact of $\alpha_0$ on the solution's profile. The background of a) was adapted from Ref. \cite{alpha_restrictions}. }
		\label{fig:solution_vs_analytical}
	\end{center}	
\end{figure}

Figure (\ref{fig:solution_vs_analytical}) summarizes the features of the stationary solutions obtained with the solver. Figure \ref{fig:solution_vs_analytical}-a) shows the regions forbidden for the value of $\alpha_0$ (grey area) that were obtained through several experiments \cite{alpha_restrictions}. In this Figure is also plotted the region of validity of the TF solutions (blueish zone), which indicates that there are a wide variety of systems, with different spatial scales, that can be described by this model. Figure \ref{fig:solution_vs_analytical}-b) shows the comparison between a numerical and the respective analytical solution, for a certain set of parameters. Both solutions are in good agreement, however, it is important to notice that the numerical solution does not assume the TF approximation, and while the exact solution assumes a compact support, the numerical one extends to infinity. This implies that there is a small fraction of the mass that is beyond $r=\mathcal{R}$. This explains some of the differences between the numerical $\rho_0$ and the theoretical prediction, which is below $4\%$. This confirms the existence of solutions that are only supported by the term proportional to the energy density (pressureless solutions) predicted in the previous section. Furthermore, it is also shown the agreement of a scaled solution from another one calculated with $\Gamma=30$, as predicted by the scaling laws in Eq. (\ref{eq:scale}). In Figure \ref{fig:solution_vs_analytical}-c) it is shown the effect of the $\alpha_0$ value in the solutions. For $\alpha_0\lesssim1\times10^{-5}$ the solutions barely change, while for large values we see an increase in the peak density while the solutions radius diminishes, since $R\propto1/(1+\alpha_0)$ for large values of $\alpha_0$. Nevertheless, we see that the numerical solutions continue to agree quite well with the TF solutions. Furthermore, the solutions plotted in \ref{fig:solution_vs_analytical}-a) are valid for a wide range of $\alpha_0$ values, and can cover the regions that are consistent with known bounds.
Finally, the results considered in this work disregard the contribution from the pressure term, but it is now possible to consider more complex situations where this term may be important. In particular, it is relevant to understand the impact of the pressure potential as well as the effect of the polytropic exponent $n$. However, a comprehensive study about the implications of this potential is beyond the scope of the present study, and we leave it for a future work.

\section{Discussion and Conclusions}

In this work we have considered a NMC curvature-matter gravity model described by functions of the scalar curvature, $f_1(R)$ and $f_2(R)$ given by Eq. (\ref{NMCfunctions}). Assuming that matter can be described as a fluid, it is found that the system admits a hydrodynamic fluid description \cite{gravity_hidro}. In this description two additional potentials arise, a Yukawa one and a potential proportional to the energy density. The latter induces a repulsive correction on the fluid equation, and the competition between the attractive potentials and the repulsive ones can, under specific conditions, give origin to stationary solutions. By transforming the hydrodynamic description into the Schrödinger-Newton-Yukawa system, we have confirmed, analytically and then through numerical methods, the existence of this new class of solutions. Furthermore, we also studied the impact of the model parameters in the structure of the solutions and shown that these are compatible with the existing bounds on the parameters of a putative new Yukawa-type interaction. 

The numerical recipe used in this work allowed us to circumvent the complexity of the field equation and to obtain a new class of solutions. To our knowledge, the existence of stable solutions without pressure constitutes a new distinct feature of this gravity model that does not exist in GR. The observational implications of this result can be fully appreciated from a more phenomenological perspective in which specific stable and stationary gravity sustainable structures are identified so that their parameters can be matched with the ones of our gravity model. This type of analysis has been previously considered, for instance, in an oceanographic context \cite{OceanExperiment} and the present work extends its range to an astrophysical setting.

Finally, the gravitational model described in this work stands as a good candidate for developing a new optical analogue. Through a detailed analysis, we have shown that, under certain approximations, the model can be described by a system of equations that is commonly found when describing light propagating in nonlinear optical systems. Thus, the next step is to search for optical materials capable of mimicking this particular Schrödinger-Newton-Yukawa model, and with these it might be possible to produce table-top experiments capable of emulating the dynamics of the gravitational model discussed in this work.

\section*{CRediT Authorship Contribuiton Statment}

Tiago D. Ferreira: Conceptualization, Software, Formal Analysis, Investigation, Writing – original draft. Nuno A. Silva: Conceptualization, Software, Writing - review $\&$ editing. Ariel Guerreiro: Conceptualization, Writing – review $\&$ editing. João Novo: Methodology, Writing – review $\&$ editing. Orfeu Bertolami - Conceptualization, Methodology, Writing – review $\&$ editing. Tiago D. Ferreira, Nuno A. Silva and Ariel Guerreiro are responsible for the numerical tools and nonlinear optics component of the work. Orfeu Bertolami and João Novo are responsible for the cosmological component of the work.

\section*{Acknowledgments}
This work is supported by the ERDF – European Regional Development Fund through the Operational Programme for Competitiveness and Internationalisation - COMPETE 2020 Programme and by National Funds through the Portuguese funding agency, FCT - Fundação para a Ciência e a Tecnologia within project $\ll$POCI-01-0145-FEDER-032257$\gg$. T.D.F. is supported by Fundação para a Ciência e a Tecnologia through Grant No. SFRH/BD/145119/2019

\bibliographystyle{apsrev4-1}



%

\end{document}